\documentclass[11pt,twoside]{article}
\usepackage{asp2014}

\aspSuppressVolSlug
\resetcounters

\begin{document}

\title{3D Reddening and Extinction Maps at the Beginning of the Gaia Era}
\author{George~A.~Gontcharov,$^1$ Aleksandr~V.~Mosenkov,$^{2,3,1}$
\affil{$^1$Central (Pulkovo) Astronomical Observatory, Russian Academy of Sciences, 
65/1 Pulkovskoye chaussee, Saint-Petersburg, 196140 Russia; \email{georgegontcharov@yahoo.com}}
\affil{$^2$Sterrenkundig Observatorium, Universiteit Gent, Krijgslaan 281, 9000 Gent, Belgium; \email{Aleksandr.Mosenkov@UGent.be}}
\affil{$^3$St. Petersburg State University, Universitetskij pr. 28, 198504 St. Petersburg, Stary Peterhof, Russia}}

\paperauthor{George~A.~Gontcharov}{georgegontcharov@yahoo.com}{orcid.org/0000-0002-6354-3884}{Central (Pulkovo) Astronomical Observatory}{Laboratory of the Dynamics of the Galaxy}{Saint-Petersburg}{}{196140}{Russia}
\paperauthor{Aleksandr~V.~Mosenkov}{Aleksandr.Mosenkov@UGent.be}{orcid.org/0000-0001-6079-7355}{Universiteit Gent}{Sterrenkundig Observatorium}{Gent}{}{}{Belgium}

\begin{abstract}
The vast majority of Gaia DR1 Tycho-Gaia astrometric solution (TGAS) stars are
located within a distance of 1200 pc from the Sun and up to 600 pc within
the Galactic plane.
Complete usage of their parallaxes is possible only if their reddening and interstellar
extinction are known.
We obtained those from a comparison of the eight most accurate and complete 3D maps
of reddening and extinction in the considered space:
the 3D map calculated from the 2D map of Schlegel et al. (1998, SFD) based on the
barometric law of the distribution of dust, 
Arenou et al. (1992), Drimmel et al. (2003),  Gontcharov (2009), Berry et al. (2012),
Gontcharov (2012), Binney et al. (2014), and Green et al. (2015).
We analyzed the statistics of the maps and their differences in the cells of
$20\times20\times20$ pc:
1) in the whole considered space,  
2) in the cells outside the dust layer only,
3) and among them, in the cells of the polar caps within 16 deg from the Galactic poles.
To a greater or lesser extent, the maps use the SFD reddening $E_{(B-V)}$ to infinity
through the whole dust layer.
As a result, they are divided into: 1) the ones which merely render the SFD reddening
inside the dust layer between the Sun (zero reddening) and layer edge (SFD reddening)
(the Drimmel et al. and Green et al. maps),
2) the ones which introduce their own reddening system only in the cells with 
robust data but follow the SFD in the rest space 
(the Berry et al. and Binney et al. maps),
and 3) the ones which are independent of the SFD map 
(the Arenou et al. and Gontcharov maps).
The comparison of the maps in the cells outside the dust layer allows us to obtain
a formula of the correction for both overestimated and minimal reddenings of the SFD map
through the dust half-layer:
$y=3.42x^3-3.62x^2+1.62x+0.04$, where y is the corrected reddening $E_{(B-V)}$ and
x is the SFD $E_{(B-V)}$ to infinity.
In this formula the minimal reddening through the dust half-layer $0.04^m$ instead of
$0.002^m$ in the SFD map was obtained
as an average of the values from the Berry, Binney and Gontcharov (2012) maps.
This value implies $E_{(B-V)}=0.06^m$ at the Galactic poles behind the dust layer 
instead of $E_{(B-V)}=0.02^m$ in the SFD map.
We point out the need for a further study of the minimal reddening through the dust
half-layer and the overestimation of higher reddenings by use of the most precise data.
Based on the data of all the maps under consideration we compile two catalogues of
the reddenings $E_{(B-V)}$:
for 630,109 cells of $20\times20\times20$ pc in the considered space 
and for 1,758,723 TGAS stars with the most accurate distances.
\end{abstract}

\section{Conclusions}

The Arenou map is based on the data for nearby stars only and has non-smooth
changes of the extinction between divided cells of the sky.
The Drimmel map is based on the models which become unreliable within few hundred pcs
from the Sun.
The Berry map refers only to a part of the sky and combines independence of the SFD
map in regions with robust data with dependence on it when the data are poor.
Gontcharov (2009) map uses a too simple model of the 3D distribution of dust and only
within the radius of 600 pc. 
The Binney map refers only to a part of the sky and shows an unexplained deviation
of the zero point in the southern polar cap.

A major source of the discrepancies between the maps is the uncertainty of the
minimal reddening through the dust half-layer. This value varies from
$0.002^m\pm0.028^m$ in the SFD to $0.040^m\pm0.016^m$ by Gontcharov (2012)  and
$0.060^m\pm0.013^m$ in Binney et al. (2014).
This uncertainty leads to the large ($0.5-1$) relative error of the reddening for the
objects at middle and high latitudes and small distances,
including type Ia supernovae and other extragalactic objects.

\acknowledgements The study was financially supported by the
``Transient and Explosive Processes in Astrophysics''
Program P-7 of the Presidium of the Russian Academy of Sciences.

\articlefigure{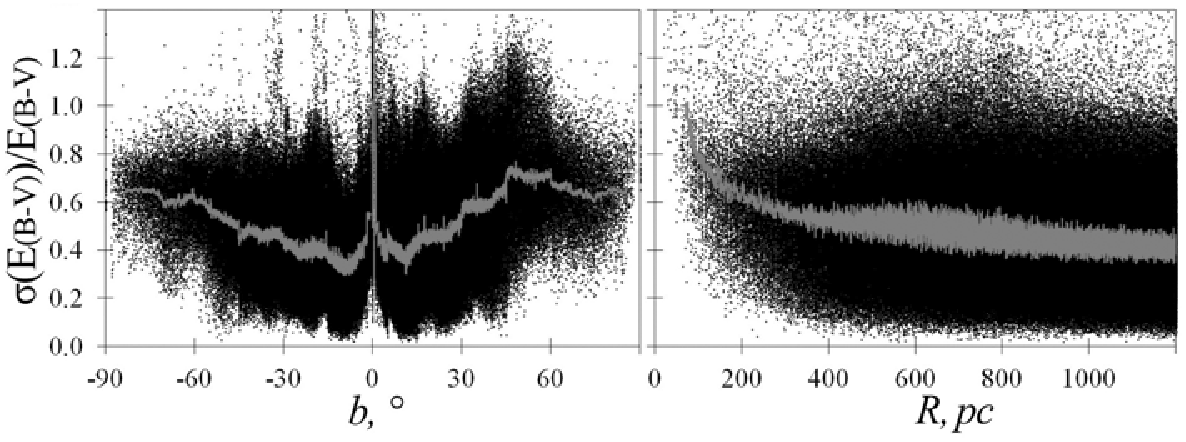}{g21}{Relative error
$\sigma(E_{(B-V)})/\overline{E_{(B-V)}}$ in 630,109 cells of the considered space
averaged over eight 3D maps in dependence on Galactic latitude and distance.
Red (grey) curve is the moving average over 222 points.
The sharp increase at $b=0$ is an artefact due to the aplication of the barometric
law to the SFD map.
The real increase of the error for middle and high latitudes and small distances
is evident.}

\end{document}